\documentclass[conference]{IEEEtran}
\IEEEoverridecommandlockouts

\usepackage{cite}

\usepackage{amssymb,amsfonts,amsthm}
\usepackage[fleqn]{amsmath}

\usepackage{algorithmic}
\usepackage{graphicx}
\usepackage{textcomp}
\usepackage{xcolor}

\usepackage{algorithm}

\usepackage{graphicx}
\usepackage{caption}
\usepackage{subcaption}
\usepackage{url}

\usepackage{balance}

\usepackage{mdframed}

\def\BibTeX{{\rm B\kern-.05em{\sc i\kern-.025em b}\kern-.08em
    T\kern-.1667em\lower.7ex\hbox{E}\kern-.125emX}}

\usepackage{todonotes}

\DeclareCaptionLabelFormat{andtable}{#1~#2  \&  \tablename~\thetable}

\begin{document}

\title{The Technological Gap Between Virtual Assistants and Recommendation Systems}

\author{\IEEEauthorblockN{Dimitrios Rafailidis}
\IEEEauthorblockA{\textit{Maastricht University} \\
Maastricht, The Netherlands \\
dimitrios.rafailidis@maastrichtuniversity.nl}
\and
\IEEEauthorblockN{Yannis Manolopoulos}
\IEEEauthorblockA{\textit{Aristotle University of Thessaloniki} \\
Thessaloniki, Greece \\
manolopo@csd.auth.gr}
}

\maketitle

\begin{abstract}
Virtual assistants, also known as intelligent conversational systems such as Google's Virtual Assistant and Apple's Siri, interact with human-like responses to users' queries and finish specific tasks. Meanwhile, existing recommendation technologies model users' evolving, diverse and multi-aspect preferences to generate recommendations in various domains/applications, aiming to improve the citizens' daily life by making suggestions. The repertoire of actions is no longer limited to the one-shot presentation of recommendation lists, which can be insufficient when the goal is to offer decision support for the user, by quickly adapting to his/her preferences through conversations. Such an interactive mechanism is currently missing from recommendation systems. This article sheds light on the gap between virtual assistants and recommendation systems in terms of different technological aspects. In particular, we try to answer the most fundamental research question, which are the missing technological factors to implement a personalized intelligent conversational agent for producing accurate recommendations while taking into account how users behave under different conditions. The goal is, instead of adapting humans to machines, to actually provide users with better recommendation services so that machines will be adapted to humans in daily life.
\end{abstract}


\section{Introduction}
Recommendation systems are intelligent agents that elicit the interests and preferences of individuals and make recommendations accordingly \cite{p1}. Recommendation systems not only have the potential to narrow down the search space of the information overload, but also to support and improve the quality of the decisions that people make in daily life. With the advent of machine learning strategies, recommendation systems can now intelligently elicit user preferences and capture their complex associations to make suggestions \cite{p2}. However, compared to existing machine learning strategies in recommendation systems, in practice there are several opportunities to elicit user information by making the underlying machine learning models more conversational and collaborative \cite{p3}. Meanwhile, recent advances in Artificial Intelligence (AI) have enabled new forms of human-computer interaction characterized by greater adaptability and better human-machine symbiosis. To facilitate the development of next generation AI agents that can truly understand and collaborate with humans, it is important that AI agents can understand and adapt to individual differences or personality traits. The AI upsurge allowed us to talk to computers via commands. Intelligent Conversational Agents (virtual assistants) have allowed us not to just talk to machines, but also accomplish our daily tasks. For example, Google Assistant, Apple's Siri, Amazon Alexa, and Microsoft Cortana have revolutionized the way we interact with phones and machines. These virtual assistants are termed as ``dialogue systems often endowed with human-like behaviour'', and they have started becoming integral parts of people's lives. Although both recommendation systems and virtual assistants are based on various machine learning strategies, there is a large technological gap between them \cite{p4, p5}. There are immense problems lying in the field of virtual assistants and recommendation systems to be solved to reach the dream we pursue, that is really adapting machines to our personal preferences while generating personalized recommendations. Existing solutions in conversational recommendation systems are either based on single round ad-hoc search engines or traditional multi-round dialog systems \cite{p6, p7, p8}, ignoring users' evolving, diverse and multi-aspect preferences when producing recommendations. The most fundamental question is:

\vspace{0.2cm}
\noindent \emph{ How can we provide people with an AI friend who will talk and give suggestions just like a human friend would have done?}
\vspace{0.2cm}

Our knowledge of bridging the gap between virtual assistants and recommendation systems is flawed. There have been many studies of virtual assistants and recommendation systems based on machine learning strategies, but no unified approach that forms a single conversational recommendation system. This article deals with the technological gap between virtual assistants and recommendation systems, shedding light on ways to develop a unified framework, not only to capture users' evolving, diverse and multi-aspect preferences, but also to consider users' interactions with the recommendation system via conversations. 

\section{Recommendation Systems}
The utility of recommendation systems cannot be overstated, given its widespread adoption in many web applications, along with its potential impact to ameliorate many problems related to over-choice. Recommendation systems provide value for people by narrowing down the set of choices and helping them explore the space of available options, or serve as a filtering component in situations of information overload. From the provider perspective, recommendation systems are personalized services that increase users' trust and loyalty, as well as obtain more knowledge about what people are really looking for. Given the explosive growth of information available on the web and Internet of Thing devices, users are often greeted with more than countless products, movies, restaurants, information about healthcare services and so on. As such, personalization is an essential strategy for facilitating user experience. Recommendation systems have been playing a vital and indispensable role in various information access systems to boost business and facilitate decision-making processes, and are pervasive across numerous web domains such as e-commerce, news and media websites. For example, 80\% of movies watched on Netflix came from recommendations \cite{p9}, 60\% of video clicks came from home page recommendation on YouTube \cite{p10}, and Amazon announced that 35\% of sales comes from recommendation systems. The core mechanisms of recommendation systems are mainly categorized into collaborative filtering, content-based recommendation systems and hybrid recommendation systems based on the types of input data. Collaborative filtering makes recommendations by learning from user-item interactions \cite{p1, p11}, content-based recommendation is based on comparisons across items' and users' auxiliary information, such as text, images and videos \cite{p12}, and hybrid models refer to recommendation systems that integrate collaborative and content-based strategies, to solve the data scarcity of user preferences and the cold-start problem with users having poor history records \cite{p13, p14, p15, p16, p17}. In a similar spirit, over the past decade recommendation algorithms for rating prediction and item ranking have steadily matured with matrix factorization and other latent factor models emerging as state-of-the-art algorithms to apply in both existing and new applications/domains \cite{p123, p124, p125, p126}. 

However, the recommendation systems algorithms are typically applied in relatively straightforward and static scenarios: given information about a user's past item preferences, can we predict whether they will like a new item or rank all unseen items based on the predicted interest? In reality, recommendation is often a more complex problem, as the evaluation of a list of recommended items never takes place in a vacuum. With richer user interaction models, more elaborate recommendation systems become possible, which can stimulate, accept and process various types of user input. At the same time the repertoire of actions is no longer limited to the one-shot presentation of recommendation lists, which can be insufficient when the goal of the system is to offer decision support for the user. State-of-the-art methods of recommendation systems are not applicable to a majority of practical scenarios due to the dynamic change of content e.g., latest news, new products, and so on. Thus, it is highly desirable to quickly adapt to users' preferences on new content through effective interactive mechanisms, such as conversations.

\section{Virtual Assistants in People's Daily Lives}
The growth of the global virtual assistants market is being primarily driven by the penetration of smartphones along with a rapid growth in the social media traffic which has led to a substantial rise in the consumer awareness about benefits offered by virtual assistants. With virtual assistants the user and system can interact for multiple semantically coherent rounds on a task through natural language dialog, and it becomes possible for the system to understand the user needs or to help users clarify their needs by asking appropriate questions to the users directly. The system has to be capable of asking aspect-based questions in the right order so as to understand the user needs, while search is conducted during the conversation, and results are provided when the system feels confident \cite{p5}. Virtual assistants interact with the user in a simplified dialogue to perform a task, support interfaces that adapt to the user's queries, and personal agents that can proactively support the user, modelling his or her needs. Nowadays, with the advent of virtual assistants, there is a proliferating demand for technology in various applications including Banking, Financial Services and Insurance, automotive, IT \& telecommunications, retail, healthcare, education and others. Recently, it has been reported that about one in six physicians in EU are already using virtual assistants \cite{p18}. It is clear that virtual assistants have now entered people's daily life to accomplish tasks.

Based on the product, the virtual assistants market has been segmented into Chatbots and smart speakers. A Chatbot is a computer program that carries out a conversation through, whereas smart speakers are a type of wireless speakers and voice command devices. Virtual assistants try to interact with human-like responses that are reasonable or interesting \cite{p3, p40, p41}. Informational virtual assistants try to help users find information or directly answer user questions. Task oriented virtual assistants try to help users finish a specific task, such as booking a flight or cancelling a trip. Virtual assistants are usually built for a specific domain, such as music, books, movies, and so on. A recent report shows how virtual assistants are currently used by their owners in the US, UK, France and Germany, with 82\% of the virtual assistants owners in these countries using virtual assistants to seek information such as news, weather, recipes, appointments, advice, offers and so on \cite{p42}. The Google Assistant is primarily available on mobile and smart home devices. The Google Assistant can engage in two-way conversations. Users primarily interact with the Google Assistant through natural voice, though keyboard input is also supported. The Google Assistant is able to search the Internet, schedule events and alarms, adjust hardware settings on the user's device, and show information from the user's Google account. Google has recently announced that the Google Assistant will be able to identify objects and gather visual information through the device's camera, and support purchasing products and sending money, as well as identifying songs. In a similar spirit, Apple's Siri is a virtual assistant which uses voice queries and a natural-language user interface to answer questions, and performs actions by delegating requests to a set of Internet services. The software adapts to users' individual language usages, searches, and preferences, with continuing use. Finally, the returned results are individualized. Alexa is a virtual assistant developed by Amazon. It is capable of voice interaction, music playback, making to-do lists, setting alarms, streaming podcasts, playing audiobooks, and providing weather, traffic, sports, and other real-time information, such as news. Microsoft Cortana is a virtual assistant that can set reminders, recognize natural voice without the requirement of keyboard input, and answer questions using information from the Bing search engine.  Microsoft recently reported that Cortana now has 133 million monthly users \cite{p43}, estimating that 325.8 million people per month will use any type of virtual assistants worldwide \cite{p44, p45}.  However, all the above virtual assistants are designed to complete certain tasks and do not capture users' personal, evolving and multi-aspect preferences.

In addition, health virtual assistants have also been designed, such as PocketSkills which supports dialectical behavioural therapy, aiming at decreasing depression and anxiety trough conversations \cite{p46}. Chatbots are usually programs that are meant to have conversations with users via text or speech methods. They are meant for specific tasks in various companies and sometimes for general chit-chat purposes. They are subset or parts of AI bots/assistants rather than being complete virtual assistants \cite{p47}. Compared to Chatbots, virtual assistants are built based on complex algorithms of Natural Language Processing (NLP), Machine Learning, and Artificial Neural Networks (ANNs), learning throughout their usage and have better performance, while Chatbots are based on fixed rules which cannot be further modified. 

With the emerging of various conversational devices, and the progress of deep learning and neural NLP research, especially on natural language dialog systems, virtual assistants based on direct user-system dialoguing has gained attention by the academia as well \cite{p35, p36, p37, p38, p39, p48, p49, p50}. Spina and Trippas \cite{p38, p39} studied the ways of presenting search results over speech-only channels and transcribing the spoken search recordings to support conversational search via deep learning, and Kang et al. \cite{p51} explored the initial and follow-up queries users tend to issue to virtual assistants. However, most of those deep learning strategies for virtual assistants focus on NLP challenges instead of recommendation systems. They neither focus on recommendation problems nor do they model and utilize users' preferences to generate recommendations via user-system conversations.

\section{The Technological Gap Between Virtual Assistants and Recommendation Systems}
Academic research in recommendation systems is largely focused on algorithmic approaches for item selection and ranking, trying to predict the ratings or generate ranked lists. However, presenting an ordered list of recommendations might not be the most suitable mechanism to support users in a decision-making problem, for example, when the user needs to clarify and refine his/her preferences. To achieve this, more interactive and possibly complex systems are required, so that users can fine-tune their profiles to provide the system with a richer repertoire of ``conversational moves''. Virtual assistants could solve this problem as users discuss with the system and enable more interactive recommendation systems without complex interfaces while at the same time providing more accurate recommendations. For example, you are considering to watch a movie but you are not sure you would enjoy it, and then you would ask your friends for advice. Alternatively, imagine that an acquaintance recommends a movie that you do not think you would enjoy. In the latter case, you would be the one willing to provide information to help your friend make better recommendations in the future. Current recommendation systems do not allow this type of interactive process to occur between the system and its users, while virtual assistants are typically oriented towards executing standalone commands rather than complex conversations. On the one hand, a plethora of personal virtual assistants have started to arise in a variety of products across domains, ranging from entertainment or retail bots to health virtual assistants. However, virtual assistants are powered by recent advances in natural language understanding and focus on conversations, not on recommendations. On the other hand, conversations in recommendation systems have to focus on balancing the explore-exploit trade-off of users.

\subsection{Information Need} 
The central difference of virtual assistants and recommendation systems is the representation of the information need: while virtual assistants, as Information Retrieval (IR) systems, typically use an explicit query prompted by the user, recommendation systems exploit user's data in an implicit manner \cite{p19}. In contrast to existing virtual assistants, recommendation systems have not only to generate accurate recommendations, but novel ones to surprise users and trigger their interest, covering users' diverse tastes and making it easier for them to understand which alternatives exist \cite{p20, p21}. 

\subsection{Scarcity of Users' Preferences} 
When virtual assistants seek information, they rely on a large amount of labelled data, which may not be available in real-world applications, such as users' preferences in recommendation systems \cite{p22}. The scarcity of users' preferences has a negative impact on the quality of recommendations of collaborative filtering models, a mainstay strategy in recommendation systems. On the contrary, virtual assistants do not account the user data scarcity. More recently, several deep learning strategies have been introduced to solve the data scarcity of users' preferences \cite{p32, p33, p34}.

\subsection{Adaptation to Evolving Preferences} 
In recommendation systems users shift their preferences over time, depending on different factors \cite{p23, p24, p25, p26, p27}. For example, curiosity leads users to explore new items contrary to their ordinary choices and/or users interact with a bias based on popularity irrespective to their history record.  Users' dynamic preferences are not yet considered by virtual assistants.

\subsection{Adjustment to Cross-domain Recommendation Tasks}
While virtual assistants are designed to complete specific tasks in users' daily life, the goal of recommendation systems is also to transfer the knowledge of users across different domains/tasks, also known as cross-domain recommendation systems. The challenge in cross-domain recommendation systems is to capture users' multi-aspect behaviours when transferring knowledge and generating recommendations for various domains \cite{p28, p29, p30, p31, p131}. Adjustment to different domains based on users' preferences is a key factor that is currently missing from virtual assistants which focus only on a specific domain.

\subsection{Transparency \& Explainability}
Another important difference between recommendation systems and virtual assistants is that in order to build trust between recommendation systems and users, it has become important to complement recommendations with explanations so that users can understand why a particular item has been suggested \cite{p13}. Transparent and explainable explanations help convincing users that the system knows them very well and makes custom-made recommendations for them. In fact, when users understand the recommendation logic, they can even be empowered to correct the system's proposals. This means that recommendations without context lack motivation for a user to pay attention to them. Adding an associated explanation for a recommendation increases user satisfaction and the persuasiveness of recommendations \cite{p118}. Nonetheless, until now, virtual assistants do not provide explanations to users.

\subsection{Preference Elicitation} 
\subsubsection{Capture users' various feedback} In recommendation systems, there are several ways to state their preferences, without involving conversations. For example, users are requested either to rate the items on a predefined scale, as well as to add comments. Other recommendation systems limit the feedback scale to ``thumbs up/down'' or positive only ``like'' statements. Users might be also requested to name a few favorite artists, movies, books, social events, and points-of-interest, to specify their interests in different categories such as ``Entertainment'', ``Politics'', ``Sports'', and so on.  While these types of user feedback is be expressed in an absolute manner, relevant studies point out that pairwise preferences are important in recommendation systems, as pairwise preferences naturally arise and are expressed by users in many decision making scenarios \cite{p107, p109}. In everyday life, there are situations where rating alternative options is not the most natural mechanism for expressing preferences and making decisions, for instance, we do not rate sweaters when we want to buy one \cite{p107}. Pairwise preferences are a pivotal issue in designing effective recommendation systems, as they can lead to larger system usability compared to absolute preferences. 

\subsubsection{Users' conversation strategies} In virtual assistants to initiate users' conversations with the system, we have to design a Conversation Manager. Users in the speaking condition  start a dialogue with the system by speaking at their computer or device, while users in the typing condition by typing into an input box. For the speaking interface, we have to support a voice-to-speech service such as \cite{p110}, to convert the audio to text. In addition, we have also to allow users to view the results and to retry or edit the results manually if the transcription results in errors, or if their microphone is not working. In general, instead of asking users to provide all requirements in one step, the Conversation Manager usually guides users through an interactive dialog, following different strategies of follow-up queries based on users' satisfaction of the results. Therefore, it is required to investigate various Conversation Manager policies to select what questions to ask and how ask, to minimize the human effort, that is the length of questioning-answering response, and emphasize on capturing user preferences based on the personality and interests.

\subsection{Quality Metrics}
Quality Information Retrieval metrics that proved successful in recommendation systems \cite{ p123, p124, p125, p126} are weak indicators to evaluate the recommendation performance of a system with conversations in real-time. The quality of conversational recommendations should be evaluated in terms of number of recommendations that a user will choose expressing the level of a participant's satisfaction; the ranking performance of the recommendation mechanism such as Normalized Discounting Cumulative Gain, Precision and Recall; the dwelling time, that is the amount of time that a participant spends before choosing a recommendation \cite{p127}. Also the goal must be to minimize the number of questions asked to obtain users' selections and consequently minimize the user time and effort. In particular, we have to determine which aspect to ask at each time with a carefully trained strategy, so that the system can always ask the most important question to improve its confidence about user needs and search results, thus keeping the conversation as short as possible, and satisfy the user needs as soon as possible. Furthermore, measuring the semantic coherence of conversations of users is also a key performance indicator. To be able to provide intelligent responses, the system must correctly model the structure and semantics of a conversation, as it is also pointed out at \cite{p128}. Thus, it is required to design numeric scores that indicate more coherent parts of a conversation and provide a signal for topic drift. For example, this will be achieved by applying the Word2Vec strategy to create the textual embedding and measure their coherence based on the respective embeddings' similarities \cite{p129}, as well as following collaborative topic modeling strategies \cite{p130}. A/B online testing  in conversational systems has also to be performed, allowing the comparison of different conversation policies \cite{p132}.

\section{Conclusion}
Summarizing, there are still many technological gaps between recommendation systems and virtual assistants that researchers have to account when designing a conversational system, that is learning users' evolving, diverse and multi-aspect preferences via human-computer conversations. The dream of having a really artificial friend to make suggestions is not far anymore. 

However, to fulfill this dream, researchers have to answer the following fundamental questions: 

\begin{itemize}
\item \emph{How can user preferences via conversations be modelled into machine learning models in recommendation systems?} 
\item \emph{Which are the right junctions to perform cross-domain recommendation with machine learning models?}
\item \emph{To what extent can we provide explainable recommendations via conversations?}
\item \emph{Which are reliable indicators to evaluate the quality of recommendation systems via conversations in real-time?} 
\end{itemize}

Filling the technological gap between recommendation systems and virtual assistants could help in building a system that is more conversational to allow users to ``work together'' (collaborate) to improve the quality of recommendations and user experience.

\bibliographystyle{unsrt}
\bibliography{IEEEfull,references}

\end{document}